\documentclass[12pt,letterpaper]{article}
  \usepackage{natbib}
  \usepackage{fullpage}

  \usepackage{amsmath}
  \usepackage{graphicx}
  \usepackage{lineno}

  \title{Unconfined Aquifer Flow Theory - from Dupuit to present}
  \author{Phoolendra K. Mishra and Kristopher L. Kuhlman}

\begin{document}

  \maketitle

\linenumbers
\section{Abstract}
Analytic and semi-analytic solution are often used by researchers and
practicioners to estimate aquifer parameters from unconfined aquifer
pumping tests. The non-linearities associated with unconfined (i.e.,
water table) aquifer tests makes their analysis more complex than
confined tests. Although analytical solutions for unconfined flow began in
the mid-1800s with Dupuit, Thiem was possibly the first to use them to
estimate aquifer parameters from pumping tests in the early 1900s. In
the 1950s, Boulton developed the first transient well test solution
specialized to unconfined flow.  By the 1970s Neuman had developed
solutions considering both primary transient storage mechanisms
(confined storage and delayed yield) without non-physical fitting
parameters. In the last decade, research into developing
unconfined aquifer test solutions has mostly focused on explicitly
coupling the aquifer with the linearized vadose zone.  Despite the
many advanced solution methods available, there still exists a need
for realism to accurately simulate real-world aquifer tests.

\section{Introduction}
Pumping tests are widely used to obtain estimates of hydraulic
parameters characterizing flow and transport processes in subsurface
(e.g., \citet{kruseman90,batu1998aquifer}). Hydraulic parameter
estimates are often used in planning or engineering applications to
predict flow and design of aquifer extraction or recharge systems.
During a typical pumping test in a horizontally extensive aquifer, a
well is pumped at constant volumetric flow rate and head observations
are made through time at one or more locations.  Pumping test data are
presented as time-drawdown or distance-drawdown curves, which are
fitted to idealized models to estimate aquifer hydraulic properties.
For unconfined aquifers properties of interest include hydraulic
conductivity, specific storage, specific yield, and possibly
unsaturated flow parameters.  When estimating aquifer properties using
pumping test drawdown data, one can use a variety of analytical
solutions involving different conceptualizations and simplifiying
assumptions.  Analytical solutions are impacted by their simplifiying
assumptions, which limit their applicability to characterize certain
types of unconfined aquifers.  This review presents the historical
evolution of the scientific and engineering thoughts concerning
groundwater flow towards a pumping well in unconfined aquifers (also
referred to variously as gravity, phreatic, or water table aquifers)
from the steady-state solutions of Dupuit to the coupled transient
saturated-unsaturated solutions.  Although it is sometimes necessary
to use gridded numerical models in highly irregular or heterogeneous
systems, here we limit our consideration to analytically derived
solutions.

\section{Early Well Test Solutions}
\subsection{Dupuit's Steady-State Finite-Domain Solutions}
\cite{dupuit1857} considered steady-state radial flow to a well
pumping at constant volumetric flowrate $Q$ [L$^3$/T] in a horizontal
homogeneous confined aquifer of thickness $b$ [L].  He used Darcy's
law \citep{darcy1856} to express the velocity of groundwater flow $u$
[L/T] in terms of radial hydraulic head gradient $\left(\partial
  h/\partial r\right)$ as
\begin{equation}
  \label{darcys-law}
  u=K\frac{\partial h}{\partial r},
\end{equation}
where $K=kg/\nu$ is hydraulic conductivity [L/T], $k$ is formation
permeability [L$^2$], $g$ is the gravitational constant [L/T$^2$],
$\nu$ is fluid kinematic viscosity [L$^2$/T], $h=\psi+z$ is hydraulic
head [L], $\psi$ is gage pressure head [L], and $z$ is elevation above
an arbitrary datum [L].  Darcy derived a form equivalent to
\eqref{darcys-law} for one-dimensional flow through sand-packed pipes.
Dupuit was the first to apply \eqref{darcys-law} to converging flow by
combining it with mass conservation $Q=\left(2\pi rb \right)u$ across
a cylindrical shell concentric with the well, leading to
\begin{equation}
  \label{dupuit_1}
  Q=K\left( 2\pi rb\right)\frac{\partial h}{\partial r}.
\end{equation}
Integrating \eqref{dupuit_1} between two radial distances $r_1$ and
$r_2$ from the pumping well, Dupuit evaluated the confined
steady-state head difference between the two points as
\begin{equation}
  \label{dupuit_confined}
  h(r_{2})-h(r_{1})=\frac{Q}{2\pi Kb}\log\left( \frac{r_2}{r_1}\right).
\end{equation}
This is the solution for flow to a well at the center of a circular
island, where a constant head condition is applied at the edge of the
island ($r_2$). 

\citet{dupuit1857} also derived a radial flow solution for unconfined
aquifers by neglecting the vertical flow component.  Following a
similar approach to confined aquifers, \citet{dupuit1857}
estimated the steady-state head difference between two distances from
the pumping well for unconfined aquifers as
\begin{equation}
  \label{dupuit_unconfined}
  h^{2}(r_{2}) - h^{2} (r_{1}) = \frac{Q}{\pi K} \log\left(
    \frac{r_2}{r_1}\right).
\end{equation}
These two solutions are only strictly valid for finite domains; when applied to
domains without a physical boundary at $r_2$, the outer radius
essentially becomes a fitting parameter.  The solutions are also used in
radially infinite systems under pseudo-static conditions, when the
shape of the water table does not change with time.

Equations \eqref{dupuit_confined} and \eqref{dupuit_unconfined}
are equivalent when $b$ in \eqref{dupuit_confined} is average head
$\left( h(r_1)+h(r_2)\right)/2$.  In developing
\eqref{dupuit_unconfined}, \cite{dupuit1857} used the following
assumptions (now commonly called the Dupuit assumptions) in context of
unconfined aquifers:
\begin{itemize}
\item the aquifer bottom is a horizontal plane;
\item groundwater flow toward the pumping wells is horizontal with no
  vertical hydraulic gradient component;
\item the horizontal component of the hydraulic gradient is constant with
  depth and equal to the water table slope; and
\item there is no seepage face at the borehole.
\end{itemize}
These assumptions are one of the main approaches to simplifying the
unconfined flow problem and making it analytically tractable.  In the
unconfined flow problem both the head and the location of the water
table are unknowns; the Dupuit assumptions eliminate one of the
unknowns.

\subsection{Historical Developments after Dupuit}
\citet{narasimhan98} and \citet{vries2007} give detailed historical
accounts of groundwater hydrology and soil mechanics; only history
relevant to well test analysis is given here.  \citet{forchheimer1886}
first recognized the Laplace equation $\nabla^2 h = 0$ governed
two-dimensional steady confined groundwater flow (to which
\eqref{dupuit_confined} is a solution), allowing analogies to be drawn
between groundwater flow and steady-state heat
conduction, including the first application of conformal mapping to
solve a groundwater flow problem.  \citet{slichter1898} also arrived
at the Laplace equation for groundwater flow, and was the first to
account for a vertical flow component.  Utilizing Dupuit's
assumptions, \citet{forchheimer1898} developed the steady-state
unconfined differential equation (to which \eqref{dupuit_unconfined}
is a solution), $\nabla^2 h^2=0$.  \citet{boussinesq1904} first gave
the transient version of the confined groundwater flow equation
$\alpha_s \nabla^2 h = \partial h/\partial t$ (where $\alpha_s=K/S_s$
is hydraulic diffusivity [L$^2$/T] and $S_s$ is specific storage
[1/L]), based upon analogy with transient heat conduction.

In Prague, \citet{thiem1906} was possibly the first to use
\eqref{dupuit_confined} for estimating $K$ from pumping tests with
multiple observation wells \citep{simmons2008}.  Equation
\eqref{dupuit_confined} (commonly called the Thiem equation) was
tested in the 1930's both in the field (\citet{wenzel1932recent}
performed a 48-hour pumping test with 80 observation wells in Grand
Island, Nebraska) and in the laboratory (\citet{wyckoff1932dupuitflow}
developed a 15-degree unconfined wedge sand tank to simulate
converging flow).  Both found the steady-state solution lacking in
ability to consistently estimate aquifer parameters.
\citet{wenzel1942} developed several complex averaging approaches (e.g., the
``limiting'' and ``gradient'' formulas) to attempt to consistently
estimate $K$ using steady-state confined equations for a finite system
from transient unconfined data.  \citet{muskat1932partialpen}
considered partial-penetration effects in steady-state flow to wells,
discussing the nature of errors associated with assumption of uniform
flux across the well screen in a partially penetrating well.  Muskat's
textbook on porous media flow \citep{muskat1937book} summarized much
of what was known in hydrology and petroleum reservoir engineering
around the time of the next major advance in well test solutions by
Theis.

\subsection{Confined Transient Flow}
\begin{figure}[htb]
  \centering
  \includegraphics[width=0.6\textwidth]{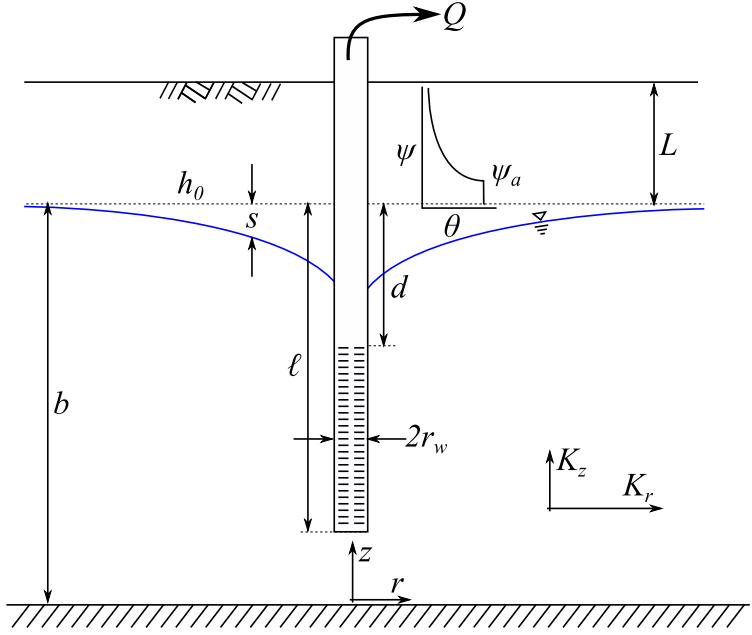}
  \caption{Unconfined well test diagram}
  \label{fig:diagram}
\end{figure}

\cite{theis1935} utilized the analogy between transient groundwater
flow and heat conduction to develop an analytical solution for
confined transient flow to a pumping well (see
Figure~\ref{fig:diagram}). He initially applied his solution to
unconfined flow, assuming instantaneous drainage due to water table
movement.  The analytical solution was based on a Green's function
heat conduction solution in an infinite axis-symmetric slab due to an
instantaneous line heat source or sink \citep{carslaw1921}.  With the
aid of mathematician Clarence Lubin, Theis extended the heat
conduction solution to a continuous source, motivated to better
explain the results of pumping tests like the 1931 test in Grand
Island.  \cite{theis1935} gave an expression for drawdown due to
pumping a well at rate $Q$ in a homogeneous, isotropic confined
aquifer of infinite radial extent as an exponential integral
\begin{equation}
  \label{theis}
  s(r,t)=\frac{Q}{4\pi T}\int_{r^2 /(4 \alpha_s t)}^{\infty}\frac{e^{-u}}{u}
  \;\mathrm{d}u,
\end{equation}
where $s=h_0(r)-h(t,r)$ is drawdown, $h_0$ is pre-test hydraulic head,
$T=Kb$ is transmissivity, and $S=S_s b$ is storativity.  Equation
\eqref{theis} is a solution to the diffusion equation, with 
zero-drawdown inital and far-field conditions,
\begin{equation}
  s(r,t=0) = s(r\rightarrow \infty,t) = 0.
\end{equation}
The pumping well was approximated by a line sink (zero radius), and
the source term assigned there was based upon \eqref{dupuit_1},
\begin{equation}{\label{boulton_sink_well}}
  \lim_{r \rightarrow 0} r\frac{\partial s}{\partial r}=-\frac{Q}{2 \pi T}.
\end{equation}

Although the transient governing equation was known through analogy
with heat conduction, the transient storage mechanism (analogous to
specific heat capacity) was not completely understood.  Unconfined
aquifer tests were known to experience slower drawdown than confined
tests, due to water supplied by dewatering the zone near the water
table, which is related to the formation specific yield (porosity less
residual water).  \citet{muskat1934transient} and
\cite{hurst1934unsteady} derived solutions to confined transient
radial flow problems for finite domains, but attributed transient
storage solely to fluid compressibility.  \citet{jacob1940} derived
the diffusion equation for groundwater flow in compressible elastic
confined aquifers, using mass conservation and Darcy's law, without
recourse to analogy with heat conduction. \citet{terzaghi1923}
developed a one-dimensional consolidation theory which only considered
the compressibility of the soil (in his case a clay), unknown at the
time to most hydrologists \citep{batu1998aquifer}.
\citet{meinzer1928} studied regional drawdown in North Dakota,
proposing the modern storage mechanism related to both aquifer
compaction and the compressiblity of water. \citet{jacob1940}
formally showed $S_s=\rho_w g(\beta_p + n\beta_w)$, where $\rho_w$ and
$\beta_w$ are fluid density [M/L$^3$] and compressibility
[LT$^2$/M], $n$ is dimensionless porosity, and $\beta_p$ is
formation bulk compressibility.  The axis-symmetric diffusion equation
in radial coordinates is
\begin{equation}
  \label{diffusion}
  \frac{\partial ^2 s}{\partial r^2}+\frac{1}{r}\frac{\partial
    s}{\partial r}=
  \frac{1}{\alpha_s}\frac{\partial s}{\partial t}.
\end{equation}

When deriving analytical expressions, the governing equation is
commonly made dimensionless to simplify presentation of results.  For
flow to a pumping well, it is convenient to use $L_C = b$ as a
characteristic length, $T_C = Sb^2/T$ as a characteristic time, and
$H_C = Q/(4 \pi T)$ as a characteristic head.  The dimensionless
diffusion equation is
\begin{equation}
  \label{diffusion}
  \frac{\partial ^2 s_D}{\partial r_D^2}+\frac{1}{r_D}\frac{\partial
    s_D}{\partial r_D}=\frac{\partial s_D}{\partial t_D},
\end{equation}
where $r_D=r/L_C$, $s_D=s/H_c$, and $t_D=t/T_C$ are scaled by
characteristic quantities.

The \cite{theis1935} solution was developed for field application to
estimate aquifer hydraulic properties, but it saw limited use because
it was difficult to compute the exponential integral for arbitrary
inputs.  \cite{wenzel1942} proposed a type-curve method that enabled
graphical application of the \cite{theis1935} solution to field data.
\cite{cooperjacob1946} suggested for large values of $t_D$ ($t_D
\geq 25$), the infinite integral in the \cite{theis1935} solution can
be approximated as
\begin{equation}
\label{JacobCooper}
s_D(t_D,r_D) = \int_{r^2/(4\alpha_st)}^{\infty} \frac{e^{-u}}{u} \; \mathrm{d}u \approx
\log_e \left(\frac{4 Tt}{r^2S}\right) - \gamma
\end{equation}
where $\gamma \approx 0.57722$ is the Euler-Mascheroni constant.  This
leads to Jacob and Cooper's straight-line simplification
\begin{equation}
  \Delta s \approx 2.3 \frac{Q}{4 \pi T}
\end{equation}
where $\Delta s$ is the drawdown over one log-cycle (base 10) of
time.  The intercept of the straight-line approximation is related to
$S$ through \eqref{JacobCooper} This approximation made estimating
hydraulic parameters much simpler at large $t_D$.  \citet{hantush1961}
later extended Theis' confined solution for partially penetrating wells.

\subsection{Observed Time-drawdown Curve}
Before the time-dependent solution of \citet{theis1935}, distance
drawdown was the diagnostic plot for aquifer test data.  Detailed
distance-drawdown plots require many observation locations (e.g., the
80 observation wells of \citet{wenzel1936TheimTest}).  Re-analyzing
results of the unconfined pumping test in Grand Island,
\cite{wenzel1942} noticed that the \cite{theis1935} solution gave
inconsistent estimates of $S_s$ and $K$, attributed to the delay in the
yield of water from storage as the water table fell.  The
\citet{theis1935} solution corresponds to the Dupuit assumptions for
unconfined flow, and can only re-create the a portion of observed
unconfined time-drawdown profiles (either late or early).  The effect
of the water table must be taken into account through a boundary
condition or source term in the governing equation to reproduce
observed behavior in unconfined pumping tests.

\begin{figure}[htb]
  \centering
  \includegraphics[width=0.8\textwidth]{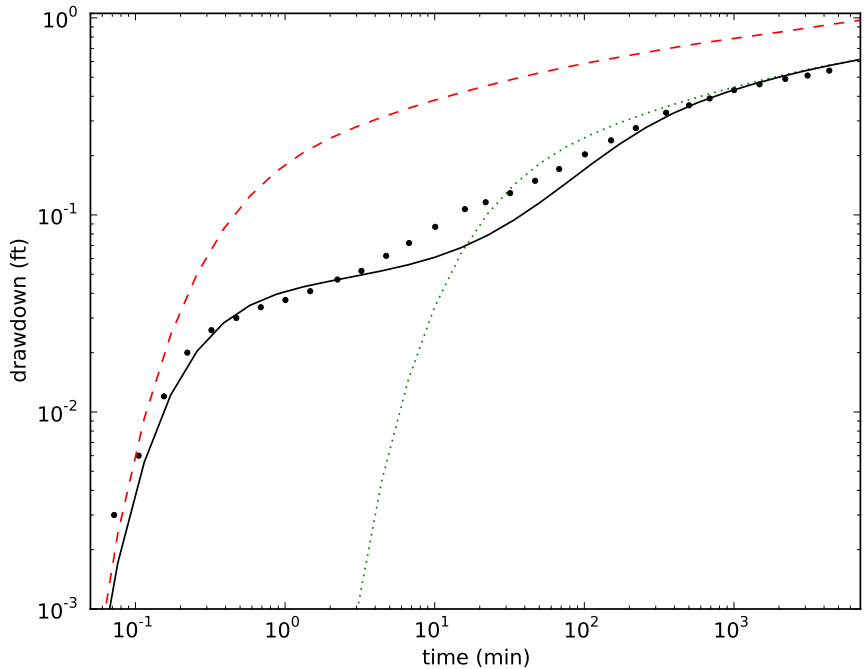}
  \caption{Drawdown data from Cape Cod \citep{moenchetal2001},
    observation well F377-037.  Upper dashed curve is
    confined model of \citet{hantush1961} with $S=S_sb$, lower dotted curve is same with
    $S=S_sb + S_y$.  Solid curve is unconfined model of
    \citet{neuman1974} using $S_y=0.23$.}
  \label{fig:capecod}
\end{figure}

\cite{walton1960} recognized three distinct segments characterizing
different release mechanisms on time-drawdown curve under water table
conditions (see Figure \ref{fig:capecod}).  A log-log time-drawdown
plot in an unconfined aquifer has a characteristic shape consisting of
a steep early-time segment, a flatter intermediate segment and a steeper
late-time segment.  The early segment behaves like the
\citet{theis1935} solution with $S=S_s b$ (water release due to bulk
medium relaxation), the late segment behaves like the
\citet{theis1935} solution with $S=S_s b + S_y$
\citep{gambolati1976transient} (water release due to water table
drop), and the intermediate segment represents a transition between
the two.  Distance-drawdown plots from unconfined aquifer tests do not
show a similar inflection or change in slope, and do not produce good
estimates of storage parameters.

\section{Early Unconfined Well Test Solutions}
\subsection{Moving Water Table Solutions Without Confined Storage}
The \cite{theis1935} solution for confined aquifers can only reproduce
either the early or late segments of the unconfined time-drawdown
curve (see Figure \ref{fig:capecod}).  \cite{boulton1954a} suggested
it is theoretically unsound to use the \citet{theis1935} solution for
unconfined flow because it does not account for vertical flow to the
pumping well.  He proposed a new mechanism for flow towards a fully
penetrating pumping well under unconfined conditions.  His formulation
assumed flow is governed by $\nabla^2 s = 0$, with transient
effects incorporated through the water table boundary condition.  He
treated the water table (where $\psi=0$, located at $z=\xi$ above the
base of the aquifer) as a moving material boundary subject to the
condition $h\left( r,z=\xi,t\right)=z$.  He considered the water table
without recharge to be comprised of a constant set of particles,
leading to the kinematic boundary condition
\begin{equation}
  \label{dynamic}
  \frac{D}{Dt}\left(h - z \right) = 0
\end{equation}
which is a statement of conservation of mass, for an incompressible
fluid.  \citet{boulton1954a} considered the Darcy velocity of the water table as
$u_z=-\frac{K_z}{S_y}\frac{\partial h}{\partial z}$ and
$u_r=-\frac{K_r}{S_y}\frac{\partial h}{\partial r}$, and expressed the
total derivative as
\begin{equation}
\label{material_derivative}
  \frac{D}{Dt}=\frac{\partial}{\partial t}-
  \frac{K_r}{S_y}\frac{\partial h}{\partial r}\frac{\partial}{\partial r}-
  \frac{K_z}{S_y}\frac{\partial h}{\partial z}\frac{\partial}{\partial z},
\end{equation}
where $K_r$ and $K_z$ are radial and vertical hydraulic
conductivity components.  Using \eqref{material_derivative}, the kinematic
boundary condition \eqref{dynamic} in terms of drawdown is
\begin{equation}{\label{free_surface}}
  \frac{\partial s}{\partial t}-\frac{K_r}{S_y}
  \left( \frac{\partial s}{\partial r} \right)^2-
  \frac{K_z}{S_y} \left( \frac{\partial s}{\partial z} \right)^2=-
  \frac{K_z}{S_y}\frac{\partial s}{\partial z}.
\end{equation}

\citet{boulton1954a} utilized the wellbore and far-field boundary
conditions of \citet{theis1935}.  He also considered the aquifer
rests on an impermeable flat horizontal boundary $\left. \partial
  h/\partial z \right|_{z=0}= 0$; this was also inferred by
\citet{theis1935} because of his two-dimensional radial flow
assumption.  \cite{dagan1967} extended Boulton's water table solution
to the partially penetrating case by replacing the wellbore
boundary condition with
\begin{equation}
  \lim_{r \rightarrow 0} r\frac{\partial s}{\partial r}=
  \begin{cases}\frac{Q}{2\pi K (\ell - d)} & b-\ell < z < b-d \\ 0 &
    \text{otherwise}\end{cases},
\end{equation}
where $\ell$ and $d$ are the upper and lower boundaries of the pumping
well screen, as measured from the initial top of the aquifer.

The two sources of non-linearity in the unconfined problem are: 1) the
boundary condition is applied at the water table, the location of
which is unknown \textit{a priori}; 2) the boundary condition applied
on the water table includes $s^2$ terms.

In order to solve this non-linear problem both Boulton and Dagan
linearized it by disregarding second order components in the
free-surface boundary condition \eqref{free_surface} and forcing the
free surface to stay at its initial position, yielding
\begin{equation}\label{boulton_linear_water_table}
  \frac{\partial s}{\partial t}=-
  \frac{K_z}{S_y}\frac{\partial s}{\partial z} \qquad\qquad z=h_0,
\end{equation}
which now has no horizontal flux component after neglecting
second-order terms.  Equation \eqref{boulton_linear_water_table} can
be written in non-dimensional form as
\begin{equation}\label{dimensionless_MWT_bc}
  \frac{\partial s_D}{\partial t_D}=-
  K_D \sigma^{\ast} \frac{\partial s_D}{\partial z_D} \qquad\qquad z_D=1,
\end{equation}
where $K_D=K_z/K_r$ is the dimensionless anisotropy ratio and
$\sigma^{\ast}=S/S_y$ is the dimensionless storage ratio.

Both \cite{boulton1954a} and \cite{dagan1967} solutions reproduce the
intermediate and late segments of the typical unconfined time-drawdown
curve, but neither of them reproduces the early segment of the curve.
Both are solutions to the Laplace equation, and therefore disregard
confined aquifer storage, causing pressure pulses to propagate
instantaneously through the saturated zone.  Both solutions exhibit an
instantaneous step-like increase in drawdown when pumping
starts.

\subsection{Delayed Yield Unconfined Response}
\cite{boulton1954b} extended Theis' transient confined theory to
include the effect of delayed yield due to movement of the water table
in unconfined aquifers.  Boulton's proposed solutions
(\citeyear{boulton1954b,boulton1963}) reproduce all three segments of
the unconfined time-drawdown curve.  In his formulation of delayed
yield, he assumed as the water table falls water is released from
storage (through drainage) gradually, rather than instantaneously as
in the free-surface solutions of \cite{boulton1954a} and
\cite{dagan1967}.  This approach yielded an integro-differential flow
equation in terms of vertically averaged drawdown $s^\ast$ as
\begin{equation}\label{boulton_solution}
  \frac{\partial ^2 s^\ast}{\partial r^2}+
  \frac{1}{r}\frac{\partial s^\ast}{\partial r}=
  \left[\frac{S}{T}\frac{\partial s^\ast}{\partial t} \right]+
  \left\lbrace\alpha S_y \int _0^t \frac{\partial s^\ast}{\partial \tau}
  e^{-\alpha (t-\tau )}\;\mathrm{d}\tau \right \rbrace
\end{equation}
which Boulton linearized by treating $T$ as constant.  The term in
square brackets is instantaneous confined storage, the term in braces
is a convolution integral representing storage released gradually
since pumping began, due to water table decline.  \cite{boulton1963}
showed the time when delayed yield effects become negligible is
approximately equal to $\frac{1}{\alpha}$, leading to the term ``delay
index''.  \cite{prickett1965} used this concept and through analysis
of large amount of field drawdown data with \cite{boulton1963}
solution, he established an empirical relationship between the delay
index and physical aquifer properties.  Prickett proposed a
methodology for estimation of $S$, $S_y$, $K$, and $\alpha$ of
unconfined aquifers by analyzing pumping tests with the
\cite{boulton1963} solution.

Although Boulton's model was able to reproduce all three segment of the
unconfined time-drawdown curve, it failed to explain the physical mechanism
of the delayed yield process because of the non-physical nature of the
``delay index'' in the \cite{boulton1963} solution.

\cite{streltsova1972a} developed an approximate solution for the
decline of the water table and $s^\ast$ in fully penetrating pumping
and observation wells.  Like \citet{boulton1954b}, she treated the
water table as a sharp material boundary, writing the two-dimensional
depth-averaged flow equation as
\begin{equation}\label{streltsova_equation}
  \frac{\partial ^2 s^\ast}{\partial r^2}+
  \frac{1}{r}\frac{\partial s^\ast}{\partial r}=
  \frac{S}{T}\left(\frac{\partial s^\ast}{\partial t}-
  \frac{\partial \xi}{\partial t} \right).
\end{equation}
The rate of water table decline was assumed to be linearly proportional to
the difference between the water table elevation $\xi$ and the
vertically averaged head $b-s^\ast$,
\begin{equation}
  \label{streltsova_watertable}
  \frac{\partial \xi}{\partial t}=\frac{K_z}{S_yb_z}
  \left( s^\ast-b+\xi \right)
\end{equation}
where $b_z=b/3$ is an effective aquifer thickness over which water
table recharge is distributed into the deep aquifer.  Equation
\eqref{streltsova_watertable} can be viewed as an approximation to the
zero-order linearized free-surface boundary condition
\eqref{boulton_linear_water_table} of \cite{boulton1954a} and
\cite{dagan1967}.  Streltsova considered the initial condition $\xi
(r,t=0)=b$ and used the same boundary condition at the pumping well
and the outer boundary $(r\rightarrow \infty )$ used by
\cite{theis1935} and \cite{boulton1963}.  Equation
\eqref{streltsova_equation} has the solution \citep{streltsova1972b}
\begin{equation}
  \label{strelstsova_sol}
  \frac{\partial \xi}{\partial t} =
  -\alpha_T \int _0^t e^{-\alpha_T (t-\tau)}
  \frac{\partial s^\ast}{\partial \tau}\;\mathrm{d}\tau
\end{equation}
where $\alpha_T = K_z/(S_yb_z)$.  Substituting \eqref{strelstsova_sol}
into \eqref{streltsova_watertable} produces solution
\eqref{boulton_solution} of \cite{boulton1954b,boulton1963}; the two
solutions are equivalent.  Boulton's delayed yield theory (like that
of Streltsova) does not account for flow in unsaturated zone but
instead treats water table as material boundary moving vertically
downward under influence of gravity.  \cite{streltsova1973} used field
data collected by \cite{meyer1962} to demonstrate unsaturated flow had
virtually no impact on the observed delayed process.  Although
Streltsova's solution related Boulton's delay index to physical
aquifer properties, it was later found to be a function of $r$
\citep{neuman1975,herrera1978}.  The delayed yield solutions of
Boulton and Streltsova do not account for vertical flow
within the unconfined aquifer through simplifying assumptions; they
cannot be extended to account for partially penetrating pumping and
observation wells.

Prickett's pumping test in the vicinity of Lawrenceville, Illinois
\citep{prickett1965} showed that specific storage in unconfined
aquifers can be much greater than typically observed values in
confined aquifers -- possibly due to entrapped air bubbles or poorly
consolidated shallow sediments. It is clear the elastic
properties of unconfined aquifers are too important to be disregarded.

\subsection{Delayed Water Table Unconfined Response}
Boulton's (\citeyear{boulton1954b,boulton1963}) models encountered
conceptual difficulty explaining the physical mechanism of water
release from storage in unconfined aquifers.  \cite{neuman1972}
presented a physically based mathematical model that treated the
unconfined aquifer as compressible (like
\citet{boulton1954b,boulton1963} and
\citet{streltsova1972a,streltsova1972b}) and the water table as a
moving material boundary (like \citet{boulton1954a} and
\citet{dagan1967}).  In Neuman's approach aquifer delayed response 
was caused by physical water table movement, he therefore proposed
to replace the phrase ``delayed yield'' by ``delayed water table
response''.

\cite{neuman1972} replaced the Laplace equation of \cite{boulton1954a}
and \cite{dagan1967} by the diffusion equation; in dimensionless
form it is
\begin{equation}
  \label{neuman_diffusion}
  \frac{\partial ^2 s_D}{\partial r_D^2}+
  \frac{1}{r_D}\frac{\partial s_D}{\partial r_D}+
  K_D\frac{\partial ^2s_D}{\partial z_D^2} =
  \frac{\partial s_D}{\partial t_D}.
\end{equation}
Like \citet{boulton1954a} and \citet{dagan1967}, Neuman treated the
water table as a moving material boundary, linearized it (using
\eqref{dimensionless_MWT_bc}), and treated the anisotropic aquifer as
three-dimensional axis-symmetric.  Like \citet{dagan1967},
\citet{neuman1974} accounted for partial penetration.  By including
confined storage in the governing equation \eqref{neuman_diffusion},
Neuman was able to reproduce all three parts of the observed
unconfined time-drawdown curve and produce parameter estimates
(including the ability to estimate $K_z$) very similar to the delayed
yield models.

Compared to the delay index models, Neuman's solution produced similar
fits to data (often underestimating $S_y$, though), but
\cite{neuman1975,neuman1979} questioned the physical nature of
Boulton's delay index.  He performed a regression fit between the
\citet{boulton1954b} and \citet{neuman1972} solutions, resulting in
the relationship
\begin{equation}
  \label{alpha-regression}
  \alpha = \frac{K_z}{S_yb}\left[3.063-0.567
    \log\left( \frac{K_Dr^2}{b^2}\right) \right]
\end{equation}
demonstrating $\alpha$ decreases linearly with $\log r$ and is
therefore not a characteristic aquifer constant.  When ignoring
the logarithmic term in \eqref{alpha-regression} the relationship
$\alpha=3K_z/(S_yb)$ proposed by \cite{streltsova1972a} is
approximately recovered.

After comparative analysis of various methods for determination of
specific yield, \cite{neuman1987} concluded the water table response
to pumping is a much faster phenomenon than drainage in the unsaturated zone
above it.

\citet{malama2011} recently proposed an alternative linearization of
\eqref{free_surface}, approximately including the effects of the
neglected second-order terms, leading to the alternative water table
boundary condition of
\begin{equation}
  \label{malama}
  S_y \frac{\partial s}{\partial t} =- K_z
  \left( \frac{\partial s}{\partial z} +
    \beta \frac{\partial^2 s}{\partial z^2} \right) \qquad z=h_0
\end{equation}
where $\beta$ is a linearization coefficient [L].  The parameter
$\beta$ provides additional adjustment of the shape of the
intermediate portion of the time-drawdown curve (beyond adjustments
possible with $K_D$ and $\sigma^\ast$ alone), leading to improved
estimates of $S_y$. When $\beta=0$ \eqref{malama} simplifies to
\eqref{boulton_linear_water_table}.

\subsection{Hybrid Water Table Boundary Condition}
The solution of \cite{neuman1972, neuman1974} was accepted by many
hydrologists ``as the preferred model ostensibly because it appears to
make the fewest simplifying assumptions'' \citep{moenchetal2001}.
Despite acceptance, \cite{nwankwor1984} and \cite{moench1995} pointed
out that significant difference might exist between measured and
model-predicted drawdowns, especially at locations near the water
table, leading to significantly underestimated $S_y$ using Neuman's
models (e.g., see Figure~\ref{fig:capecod}).  \citet{moench1995}
attributed the inability of Neuman's models to give reasonable
estimates of $S_y$ and capture this observed behavior near the water
table due to the later's disregard of ``gradual drainage''.  In an
attempt to resolve this problem, \cite{moench1995} replaced the
instantaneous moving water table boundary condition used by Neuman
with one containing a \citet{boulton1954b} delayed yield convolution
integral,
\begin{equation}
\label{moench_hybrid}
  \int _0^t\frac{\partial s}{\partial \tau}
  \sum _{m=1}^M \alpha _m e^{-\alpha _m (t-\tau )}\;\mathrm{d}\tau =-
  \frac{K_z}{S_y}\frac{\partial s}{\partial z}
\end{equation}
This hybrid boundary condition ($M=1$ in \citet{moench1995}) included
the convolution source term \citet{boulton1954b,boulton1963} and
\citet{streltsova1972a,streltsova1972b} used in their depth-averaged
governing flow equations.  In addition to this new boundary condition,
\cite{moench1995} included a finite radius pumping well with wellbore
storage, conceptually similar to how \citet{papadopulosandcooper1967}
modified the solution of \citet{theis1935}.  In all other respects,
his definition of the problem was similar to \cite{neuman1974}.

Moench's solution resulted in improved fits to experimental data and
produced more realistic estimates of specific yield
\citep{moenchetal2001}, including the use of multiple delay parameters
$\alpha_m$ \citep{moench2003}.  \citet{moenchetal2001} used
\eqref{moench_hybrid} with $M = 3$ to estimate hydraulic parameters in
the unconfined aquifer at Cape Cod.  They showed that $M=3$ enabled a
better fit to the observed drawdown data than obtained by $M<3$ or the
model of \cite{neuman1974}.  Similar to the parameter $\alpha $ in
Boulton's model, the physical meaning of $\alpha _m$ is not clear.

\section{Unconfined Solutions Considering Unsaturated Flow}
As an alternative to linearizing the water table condition of
\citet{boulton1954a}, the unsaturated zone can be explicitly included.
The non-linearity of unsaturated flow is substituted for the
non-linearity of \eqref{free_surface}.  By considering the vadose
zone, the water table is internal to the domain, rather than a
boundary condition.  The model-data misfit in Figure~\ref{fig:capecod}
at ``late intermediate'' time is one of the motivations for
considering the mechanisms of delayed yield and the effects of the
unsaturated zone.

\subsection{Unsaturated Flow Without Confined Aquifer Storage}
\cite{kroszynskidagan1975} were the first to account analytically for
the effect of the unsaturated zone on aquifer drawdown.  They extended
the solution of \cite{dagan1967} by accounting for unsaturated flow
above the water table.  They used Richards' equation for
axis-symmetric unsaturated flow in a vadose zone of thickness $L$
\begin{equation}
  \label{richards}
  K_r\frac{1}{r}\frac{\partial}{\partial r}\left( k(\psi
    )r\frac{\partial \sigma}{\partial r}\right)+
  K_z\frac{\partial}{\partial z}\left( k(\psi )\frac{\partial
      \sigma}{\partial z}\right) = C(\psi)\frac{\partial \sigma
  }{\partial t}
  \qquad
  \xi < z <b+L
\end{equation}
where $\sigma = b + \psi_a - h$ is unsaturated zone drawdown [L],
$\psi _a$ is air-entry pressure head [L], $0 \leq k(\psi )\leq 1$ is
dimensionless relative hydraulic conductivity, $C(\psi)=d\theta/d
\psi$ is the moisture retention curve [1/L], and $\theta$ is dimensionless
volumetric water content (see inset in Figure~\ref{fig:diagram}).
They assumed flow in the underlying saturated zone was governed by the
Laplace equation (like \citet{dagan1967}).  The saturated and
unsaturated flow equations were coupled through interface conditions
at the water table expressing continuity of hydraulic heads and normal
groundwater fluxes,
\begin{eqnarray}
  s=\sigma\qquad
  \nabla s \cdot \textbf{n}=\nabla \sigma \cdot \textbf{n} \qquad
  z= \xi
\end{eqnarray}
where $\textbf{n}$ is the unit vector perpendicular to the water
table.

To solve the unsaturated flow equation \eqref{richards},
\citet{kroszynskidagan1975} linearized \eqref{richards} by adopting
the \cite{gardner1958} exponential model for the relative hydraulic
conductivity, $k(\psi )=e^{\kappa_a (\psi -\psi_a)}$, where $\kappa_a$
is the sorptive number [1/L] (related to pore size).  They adopted the
same exponential form for the moisture capacity model, $\theta (\psi
)=e^{\kappa_k (\psi -\psi_k)}$, where $\psi_k$ is the pressure at
which $k(\psi)=1$, $\kappa_a=\kappa_k$, and $\psi_a=\psi_k$, leading
to the simplified form $C(\psi)=S_y\kappa_a e^{\kappa_a (\psi
  -\psi_a)}$.  In the limit as $\kappa_k=\kappa_a \rightarrow \infty$
their solution reduces to that of \cite{dagan1967}.  The relationship
between pressure head and water content is a step function.
\cite{kroszynskidagan1975} took unsaturated flow above the water table
into account but ignored the effects of confined aquifer storage,
leading to early-time step-change behavior similar to
\citet{boulton1954a} and \citet{dagan1967}.

\subsection{Increasingly Realistic Saturated-Unsaturated Well Test Models}
\cite{mathiasbutler2006} combined the confined aquifer flow equation
\eqref{neuman_diffusion} with a one-dimensional linearized version of
\eqref{richards} for a vadose zone of finite thickness.  Their water
table was treated as a fixed boundary with known flow conditions,
decoupling the unsaturated and saturated solutions at the water table.
Although they only considered a one-dimensional unsaturated zone, they
included the additional flexibility provided by different exponents
$(\kappa_a \neq \kappa_k)$.  \cite{mathiasbutler2006} did not consider a partially
penetrating well, but they did note the possibility of accounting for
it in principle by incorporating their uncoupled drainage function in
the solution of \citet{moench1997}, which considers a partially
penetrating well of finite radius.

\cite{tartakovskyneuman2007} similarly combined the confined aquifer
flow equation \eqref{neuman_diffusion}, but with the original
axis-symmetric form of \eqref{richards} considered by
\citet{kroszynskidagan1975}.  Also like \citet{kroszynskidagan1975},
their unsaturated zone was characterized by a single exponent
$\kappa_a=\kappa_k$ and reference pressure head
$\psi_a=\psi_k$. Unlike \citet{kroszynskidagan1975} and
\citet{mathiasbutler2006}, \cite{tartakovskyneuman2007} assumed an
infinitely thick unsaturated zone.

\cite{tartakovskyneuman2007} demonstrated flow in the unsaturated zone
is not strictly vertical.  Numerical simulations by \citet{moench2008}
showed groundwater movement in the capillary fringe is more horizontal
than vertical.  \cite{mathiasbutler2006} and \citet{moench2008} showed
that using the same exponents and reference pressure heads for
effective saturation and relative permeability decreases model
flexibility and underestimates $S_y$.  \citet{moench2008} predicted an
extended form of \citet{tartakovskyneuman2007} with two separate
exponents, a finite unsaturated zone, and wellbore storage would
likely produce more physically realistic estimates of $S_y$.

\cite{mishraneuman2010} developed a new generalization of the solution
of \cite{tartakovskyneuman2007} that characterized relative hydraulic
conductivity and water content using $\kappa_a \neq \kappa_k$, $\psi_a
\neq \psi_k$ and a finitely thick unsaturated zone.
\citet{mishraneuman2010} validated their solution against numerical
simulations of drawdown in a synthetic aquifer with unsaturated
properties given by the model of \cite{vangenuchten1980}.  They also
estimated aquifer parameters from Cape Cod drawdown data
\citep{moenchetal2001}, comparing estimated \citet{vangenuchten1980}
parameters with laboratory values \citep{maceetal1998}.

\cite{mishraneuman2011} further extended their
\citeyear{mishraneuman2010} solution to include a finite-diameter
pumping well with storage.  \citet{mishraneuman2010,mishraneuman2011}
were the first to estimate non-exponential model unsaturated aquifer
properties from pumping test data, by curve-fitting the exponential
model to the \cite{vangenuchten1980} model.  Analyzing pumping test
data of \citet{moenchetal2001} (Cape Cod, Massachusetts) and
\citet{nwankwor1984,nwankwor1992} (Borden, Canada), they estimated
unsaturated flow parameters similar to laboratory-estimated values for
the same soils.

\section{Future Challenges}
The conceptualization of groundwater flow during unconfined pumping
tests has been a challenging task that has spurred substantial
theoretical research in the field hydrogeology for decades.
Unconfined flow to a well is non-linear in multiple ways, and the
application of analytical solutions has required utilization of
advanced mathematical tools.  There are still many additional
challenges to be addressed related to unconfined aquifer pumping
tests, including:
\begin{itemize}
\item Hysteretic effects of unsaturated flow. Different exponents
  and reference pressures are needed during drainage and recharge
  events, complicating simple superposition needed to handle multiple
  pumping wells, variable pumping rates, or analysis of recovery data.
\item Collecting different data types.  Validation of existing models
  and motivating development of more realistic ones depends on more
  than just saturated zone head data. Other data types include vadose
  zone water content \citep{meyer1962}, and hydrogeophysical data like
  microgravity \citep{damiata2006} or streaming potentials
  \citep{malama2009}.
\item Moving water table position.  All solutions since
  \citet{boulton1954a} assume the water table is fixed horizontal
  $\xi(r,t)=h_0$ during the entire test, even close to the pumping
  well where large drawdown is often observed.  Iterative numerical
  solutions can accommodate this, but this has not been included in an
  analytical solution.
\item Physically realistic partial penetration. Well test solutions
  suffer from the complication related to the unknown distribution of
  flux across the well screen.  Commonly, the flux distribution is
  simply assumed constant, but it is known that flux will be higher
  near the ends of the screened interval that are not coincident with
  the aquifer boundaries.
\item Dynamic water table boundary condition.  A large increase in
  complexity comes from explicitly including unsaturated flow in
  unconfined solutions.  The kinematic boundary condition expresses
  mass conservation due to water table decline.  Including an
  analogous dynamic boundary condition based on a force balance
  (capillarity vs. gravity) may include sufficient effects of
  unsaturated flow, without the complexity associated with the
  complete unsaturated zone solution.
\item Heterogeneity.  In real-world tests heterogeneity is present at
  multiple scales.  Large-scale heterogeneity (e.g., faults or rivers)
  can sometimes be accounted in analytical solutions using the method
  of images, or other types of superpostion.  A stochastic approach
  \citep{neuman2004type} could alternatively be developed to estimate
  random unconfined aquifer parameter distribution parameters.
\end{itemize}

Despite advances in considering physically realistic unconfined flow,
most real-world unconfined tests (e.g., \citet{wenzel1942},
\citet{nwankwor1984,nwankwor1992}, or \citet{moenchetal2001}) exhibit
non-classical behavior that deviates from the early-intermediate-late
behavior predicted by the models summarized here. We must continue to
strive to include physically relevant processes and representatively
linearize non-linear phenomena, to better understand, simulate and
predict unconfined flow processes.

\section*{Acknowledgments}
This research was partially funded by the Environmental Programs
Directorate of the Los Alamos National Laboratory. Los Alamos National
Laboratory is a multi-program laboratory managed and operated by Los
Alamos National Security (LANS) Inc. for the U.S. Department of
Energy's National Nuclear Security Administration under contract
DE-AC52-06NA25396.

Sandia National Laboratories is a multi-program laboratory managed and
operated by Sandia Corporation, a wholly owned subsidiary of Lockheed
Martin Corporation, for the U.S. Department of Energy's National
Nuclear Security Administration under contract DE-AC04-94AL85000.



\bibliography{review_paper}
\bibliographystyle{abbrvnat}






\end{document}